\documentclass[a4paper]{jpconf}
\usepackage{graphicx}

\newcommand{\beq}{\begin{equation}}
\newcommand{\eeq}{\end{equation}}
\newcommand{\bea}{\begin{eqnarray}}
\newcommand{\eea}{\end{eqnarray}}

\begin{document}
\title{Advances in non-relativistic matter via complex Langevin approaches}

\author{Joaqu{\'i}n E. Drut}
\address{Department of Physics and Astronomy, University of North Carolina,
Chapel Hill, North Carolina, 27599-3255, USA}
\ead{drut@email.unc.edu}

\begin{abstract}
The recent progress in understanding the mathematics 
of complex stochastic quantization, as well as its application to 
quantum chromodynamics in situations that have a complex phase problem (e.g. finite quark density, real time), 
has opened up an intriguing possibility for non-relativistic many-body physics which has so far remained 
largely unexplored. In this brief contribution, I review a few specific examples of advances in the characterization 
of the thermodynamics of non-relativistic matter in a variety of one-dimensional
cases affected by the sign problem: repulsive interactions, finite polarization, finite mass imbalance, and 
projection to finite systems to obtain virial coefficients.
\end{abstract}
\section{Introduction}

This contribution is concerned with computational approaches to the generic quantum many-body problem, 
as it appears in high-energy, nuclear, condensed matter, and atomic physics; and quantum chemistry.
The standard non-perturbative computational methods to this problem can be 
roughly divided into two large sets: direct methods and stochastic methods [which we will
collectively refer to below as quantum Monte Carlo methods (QMC)]. 
In the former category we find exact diagonalization and configuration interaction approaches. While 
these are very appealing due to their potential for high precision, their memory requirements
(especially in 3 spatial dimensions) can quickly become prohibitive as the system size is increased.
Stochastic methods, which are the focus of this work, tend to have much milder memory requirements 
but suffer from other challenging issues. 

Indeed, by far, most interesting problems in quantum many-body physics suffer from the infamous
sign problem when one attempts to compute, using stochastic methods, essentially any useful quantity,
be it ground-state properties, thermodynamics, or dynamic response. The sign problem is a severe 
signal-to-noise issue characterized by an exponentially decreasing signal-to-noise ratio as the volume of 
spacetime is increased. The hallmark of the sign problem can often be detected without doing a single 
calculation: the probability measure in the path integral may be real but of indefinite sign, or it may be complex,
in which case we speak of a `phase problem'.

In spite of the general severity of the problem, in some cases it is possible to carry out calculations using a simple 
re-weighting technique. Starting from the path integral formulation of the quantum many-body problem, the expectation 
value of an operator $\hat O$ will generally assume the schematic form
\beq
\langle \hat O\rangle = \frac{\sum_{\sigma} P[\sigma] O[\sigma]}{\sum_{\sigma} P[\sigma]},
\eeq
where $P[\sigma]$ is what one would normally want to identify as a probability measure (usually a fermion 
determinant and other pieces of the action), but which will typically be complex. Rearranging the above expression 
by separating the magnitude $|P[\sigma]|$ and the phase $e^{i\phi[\sigma]}$, and multiplying and dividing by a 
normalization factor, one obtains
\beq
\langle \hat O\rangle = \frac{\sum_{\sigma} |P[\sigma]| e^{i\phi[\sigma]} O[\sigma]}{\sum_{\sigma} |P[\sigma]|} 
\frac{\sum_{\sigma} |P[\sigma]|}{\sum_{\sigma} |P[\sigma]| e^{i\phi[\sigma]}} 
= \frac{\langle \langle e^{i\phi[\sigma]} O[\sigma] \rangle \rangle}{\langle \langle e^{i\phi[\sigma]} \rangle\rangle}.
\eeq
In this expression, the double angle bracket denotes an expectation value with respect to the non-negative 
measure $|P[\sigma]|$. One may thus use $|P[\sigma]|$ as the probability in conventional Metropolis-based 
methods. However, the price of such a re-weighting trick can be high: both the numerator and the denominator in the above 
expression will approach zero exponentially (as a function of the system size), quickly going beyond the precision of any reasonable 
calculation (unless astronomical numbers of samples are used), such that obtaining their ratio becomes impractical.

In spite of remarkable advances toward taming this issue (i.e. find ways to make the phase average stay as close to 1 as possible), 
the search for a more generic or practical solution beyond the above re-weighting idea continues.
Remarkably, some lines of research have shown that in specific cases the sign problem can be completely avoided 
(see e.g.~\cite{SignProblemReview1, SignProblemReview2}); however, we will here focus on methods that attempt to overcome 
the problem by implementing
some form of complex stochastic quantization, also known as complex Langevin (CL), which was brought up in Refs.~\cite{Parisi1983, Klauder1984}.
The idea of complex stochastic quantization is, in fact, a line of research that has flourished in recent years after multiple abandoned 
attempts in the area of lattice QCD~\cite{FiniteDQCD0,FiniteDQCD3}. As it appears that the understanding of CL is now 
far better than ever before, it is an appealing candidate for addressing the sign problem in areas outside QCD.

In this contribution we show some of the recent attempts to use CL to overcome the sign problem in a class
of non-relativistic many-body problems, keeping in mind their realization in ultracold atom experiments. 
We begin by describing some of the operational aspects of the method, leaving the formal aspects to
the excellent extant literature (see e.g.~\cite{CLSufficientConditions00,CLSufficientConditions01,CLSufficientConditions02,CLSufficientConditions03,CLSufficientConditions04}). 
We then review recent results for spin-$1/2$ fermions in one spatial dimension (1D)
focusing on repulsive interactions at finite temperature, spin polarized systems, and mass-imbalanced systems.
Finally, we show progress towards implementing particle projection algorithms via CL, which enables the 
stochastic calculation of high-order virial coefficients. Note that, in the 1D cases described here, we implement contact
interactions, as befits dilute gases. In some cases the sign problem can be avoided entirely by using other algorithms~\cite{ZhangPC}; 
however, those algorithms do not generalize to higher dimensions, which makes them less interesting for our objectives.

\section{Outline of the methods}

For reference, it is useful to recall here the basic operations involved in the hybrid Monte Carlo (HMC) 
algorithm~\cite{HMC1,HMC2,MCReview,HSLee2}.
We begin with the path integral form of the partition function
\beq
\mathcal{Z} = \int \mathcal{D}\sigma \, e^{-S[\sigma]},
\eeq
where the effective action $S[\sigma]$ is typically the logarithm of a fermion determinant and potentially includes
pure $\sigma$ terms, as in QCD (though not necessarily so in non-relativistic physics).

When $S[\sigma]$ is real, the usual methods sample the field $\sigma$ according to $P[\sigma] = e^{-S[\sigma]}$. The HMC algorithm
carries out that sampling in a global fashion (i.e. modifying the whole $\sigma$ field at the same time) by
enlarging the phase space with the introduction of an auxiliary momentum field $\pi$ conjugate to $\sigma$
whose dynamics factorizes in the partition function and therefore does not alter the physics of the problem. Explicitly,
\beq
\mathcal{Z} = \int \mathcal{D}\sigma  \mathcal{D}\pi \, e^{-\mathcal H[\sigma,\pi]},
\eeq
where $\mathcal H[\sigma,\pi] = \frac{1}{2}\sum_{x} \pi(x)^2 + S[\sigma]$. 
The classical equations of motion for $\mathcal H[\sigma,\pi]$, in a fictitious phase-space time $t$,
are then used to update the fields: 
\bea
\dot \sigma &=& \pi, \\
\dot \pi &=& -\frac{\delta S[\sigma]}{\delta \sigma}.
\eea
In the above phase-space evolution, a trajectory of length $t \sim O(1)$ is considered a full update of the field (although
several such trajectories are usually required to ensure decorrelation). In between trajectories, a Metropolis step
is implemented to ensure that the correct distribution is being sampled, and the
momentum $\pi$ is refreshed using a Gaussian distribution. Note that, in such a classical evolution,
the fictitious energy given by the value of $\mathcal H$ is conserved (as long as the integrator is accurate enough), which ensures a 
very high acceptance rate in the Metropolis step. The HMC algorithm thus succeeds at implementing global updates with
high acceptance rate. From this brief operational description, it can be inferred that the most costly part of the algorithm
is the calculation of the force ${\delta S[\sigma]}/{\delta \sigma}$.

The stochastic quantization sister of the HMC method is what one may call real Langevin (RL)~\cite{RL2,RL3}, which enables global updates as well. In RL, however, there is no Metropolis step and the equations of motion take the form
\bea
\dot \sigma = -\frac{\delta S[\sigma]}{\delta \sigma} + \eta,
\eea
where we note that there is no auxiliary momentum field $\pi$ but a $t$-dependent noise field $\eta$ appears instead.
The latter satisfies $\langle \eta(x,\tau) \rangle = 0$ and $\langle \eta(x,\tau) \eta(x',\tau') \rangle = 2 \delta_{x,x'}\delta_{\tau,\tau'}$
for spacetime points $(x,\tau)$ and $(x',\tau')$, and may be chosen to be Gaussian. The similarities between RL and HMC are clear; most notably, the 
calculation of the `drift' (as is often called in the context of stochastic quantization) given by ${\delta S[\sigma]}/{\delta \sigma}$,
is the most computationally intensive part of the method.

The conventional mathematical underpinnings of HMC and RL depend on $P[\sigma]$ being positive semidefinite
(i.e. $S[\sigma]$ being real), a property which fails to hold in non-relativistic physics for repulsive interactions (away from the half-filling point), polarized systems, 
mass-imbalanced systems, and so on, and the calculation is then said to have a sign problem (or more generally a complex-phase problem). In the case of HMC, this means that the Metropolis step is simply no longer well-defined and thus the algorithm is no longer available. 
For RL, on the other hand, a generalization is possible into what we referred to above as CL.

Operationally, in CL one complexifies the HS field $\sigma$ via
\beq
\sigma = \sigma_R + i \sigma_I,
\eeq
where $\sigma_R$ and $\sigma_I$ are both real fields and defines equations of motion by
\bea
\delta \sigma_R &=& -\textrm{Re}\left[\frac{\delta S[\sigma]}{\delta \sigma}\right] \delta t + \eta \sqrt{\delta t}, \\
\delta \sigma_I &=& -\textrm{Im}\left[\frac{\delta S[\sigma]}{\delta \sigma}\right] \delta t,
\eea
where now $S[\sigma]$ is to be understood as a complex function of the complex variable $\sigma$.
Clearly, when the action is real, the imaginary part of the force vanishes and CL reduces to RL.
We also note that we chose real noise, but complex noise is also a possibility (see e.g. Ref.~\cite{CLSufficientConditions00}).

\section{Results}

As anticipated, this section reviews some of the recent applications of the complex 
Langevin method to non-relativistic systems. Four different situations are briefly described (all corresponding to
fermions in 1D with a zero-range interaction): repulsive interactions, finite polarization, finite mass imbalance, 
and particle number projection. 
It should be pointed out that, while 1D systems can be tackled with the Bethe ansatz (see e.g. Ref.~\cite{BA}), although the latter 
runs into difficulties at finite temperature and in the grand-canonical ensemble, and at finite mass imbalance.

\subsection{Spin-$1/2$ fermions in 1D: repulsive interactions at finite-temperature}

\begin{figure}[h]
  \begin{center}
   \includegraphics[scale=0.62]{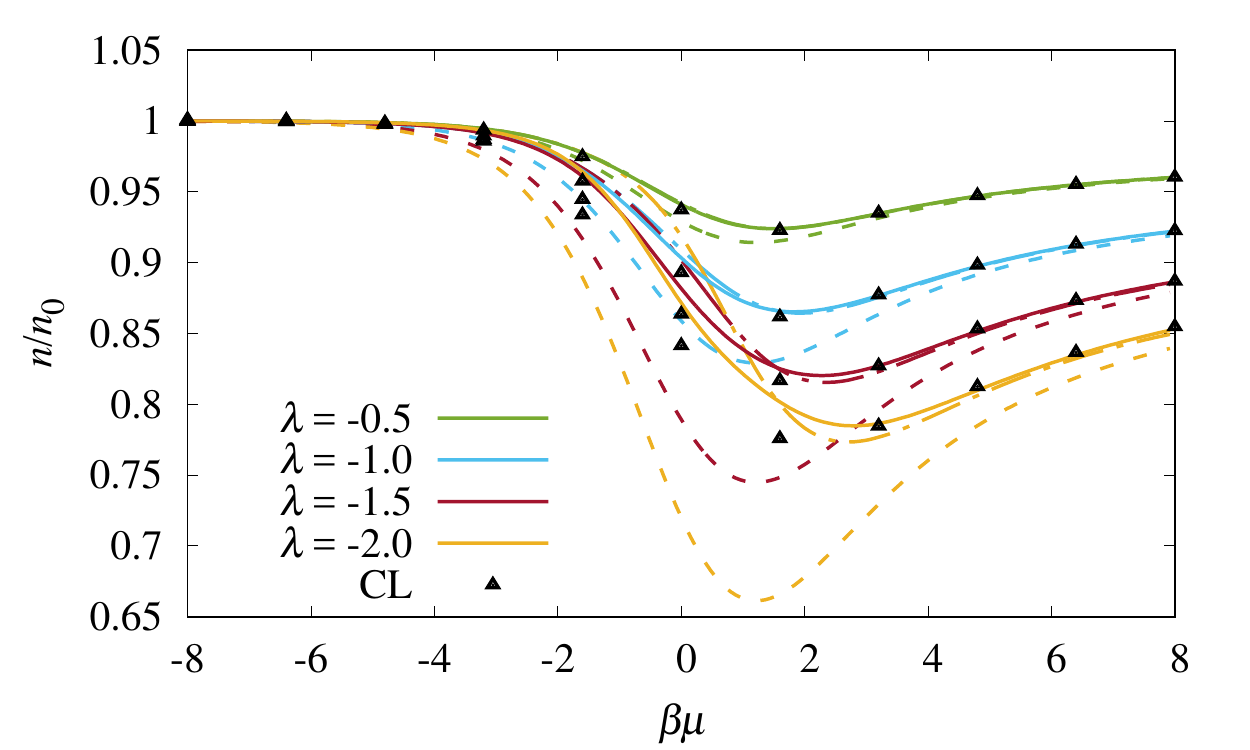}
   \includegraphics[scale=0.52]{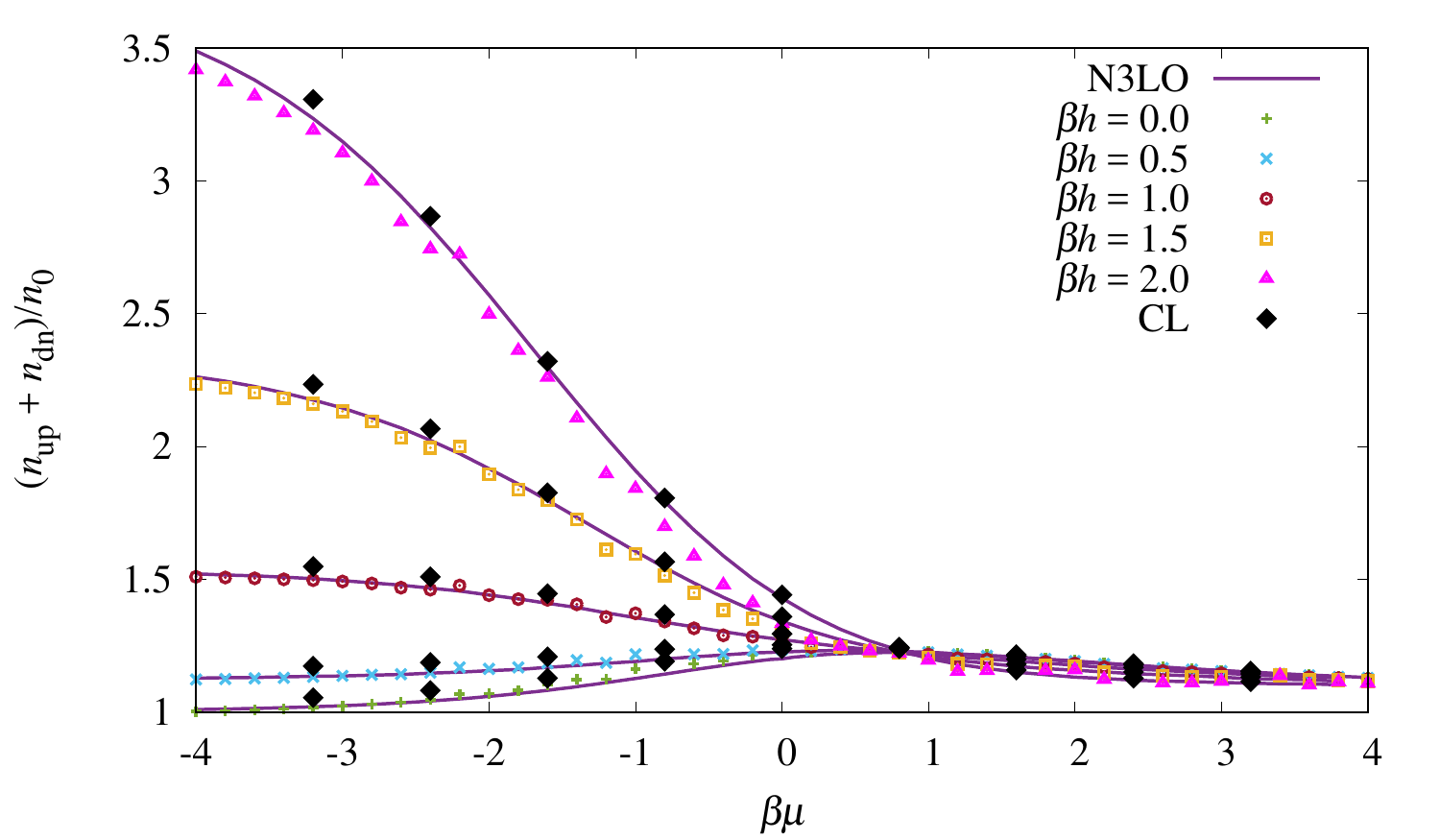}
  \end{center}
  \caption{Left panel: Density equation of state of 1D fermions with repulsive interactions. Black diamonds show the CL results.
  Dashed, dash-dotted, and solid lines show results of NLO, N2LO, and N3LO lattice perturbation theory,   respectively~\cite{PRDAndrew}.
  Right panel: Density equation of state of 1D fermions with attractive interactions at finite chemical potential asymmetry
  $\beta h$. Black diamonds show the CL results.
  Solid lines show results of N3LO lattice perturbation theory.
  Colored data points show the results of imaginary-polarization calculations of Ref.~\cite{iHMCAndrew}.
  }
  \label{Fig:RepulsiveEOS}
\end{figure}

Our first results with CL focused on the density equation of state of unpolarized fermions with 
repulsive interactions at finite temperature, which were first published in Ref.~\cite{PRDAndrew}. 
The auxiliary field method gives a sign problem for this system regardless of the dimension, 
but we focused on 1D as a starting point and compared our CL results with
lattice perturbation theory up to next-to-leading (NLO), next-to-next-to-leading (N2LO), and
next-to-next-to-next-to-leading order (N3LO), in Fig.~\ref{Fig:RepulsiveEOS} (left panel)
for several values of the coupling $\lambda$.

In the region of large fugacity, i.e. $\beta \mu \gg 0$, where we expect perturbative results to
be reliable, we find that CL agrees with them. As the coupling is increased, perturbation theory 
begins to break down, but CL continues to converge, giving a prediction for the grand canonical
equation of state.
%

\subsection{Spin-$1/2$ fermions in 1D: finite polarization at finite-temperature}

Following up on the success of our study of repulsive interactions, we move on to polarized systems,
which are especially interesting due to the possibility of realizing exotic superfluids such as the 
Fulde-Ferrell-Larkin-Ovchinnikov phase. As in the previous case, we compared with perturbative results, but also with
the imaginary-polarization approach of Ref.~\cite{iHMCAndrew} (which used an idea first put forward
in Ref.~\cite{iHMCidea}), in Fig.~\ref{Fig:RepulsiveEOS} (right panel).
This case is more remarkable than the previous one because we obtain results that are not only in agreement
with N3LO perturbation theory but also with the entirely different non-perturbative approach of
imaginary polarization.

\subsection{Spin-$1/2$ fermions in 1D: ground-state of mass-imbalanced systems}

\begin{figure}[h]
  \begin{center}
  \includegraphics[scale=0.35]{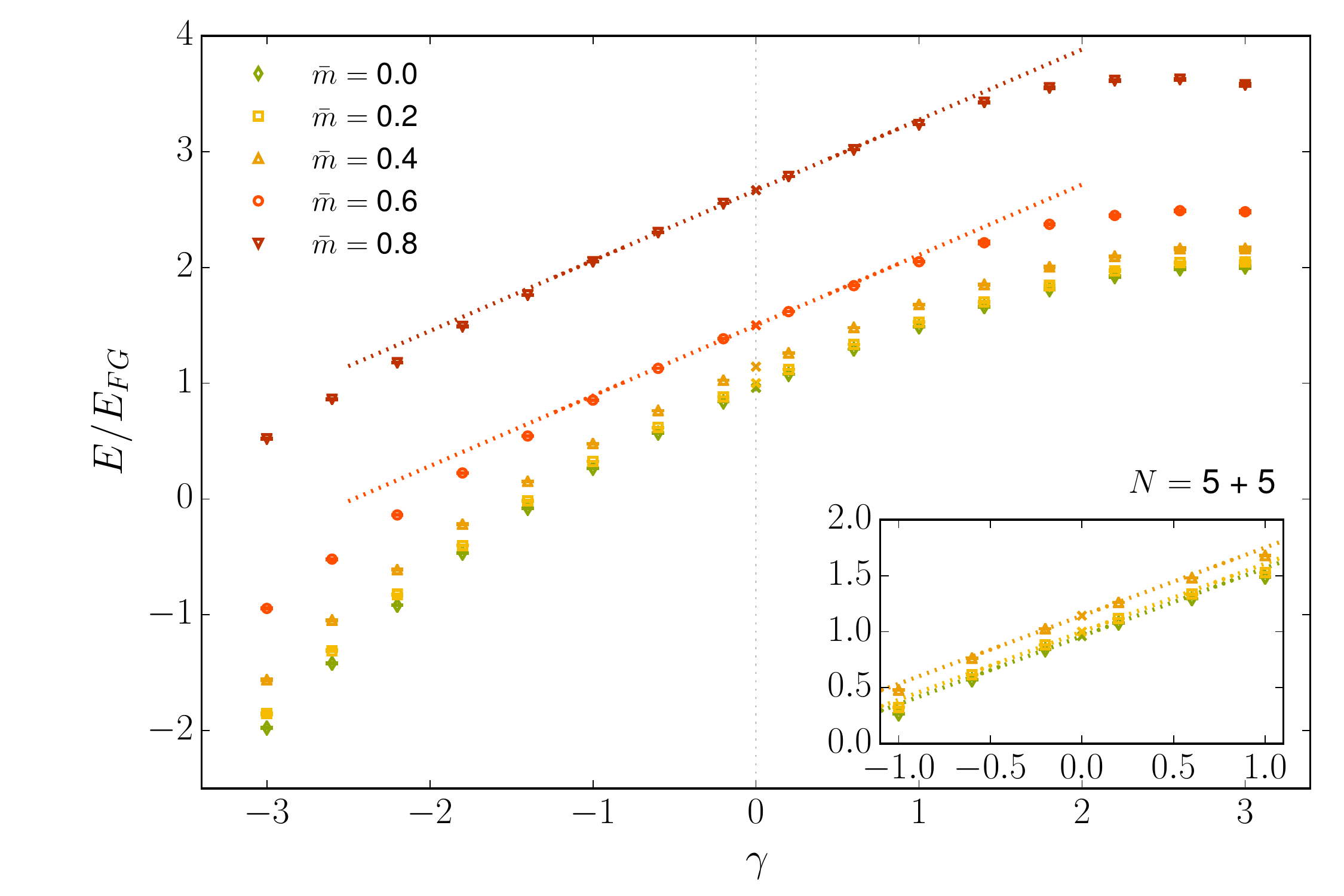}
  \includegraphics[scale=0.62]{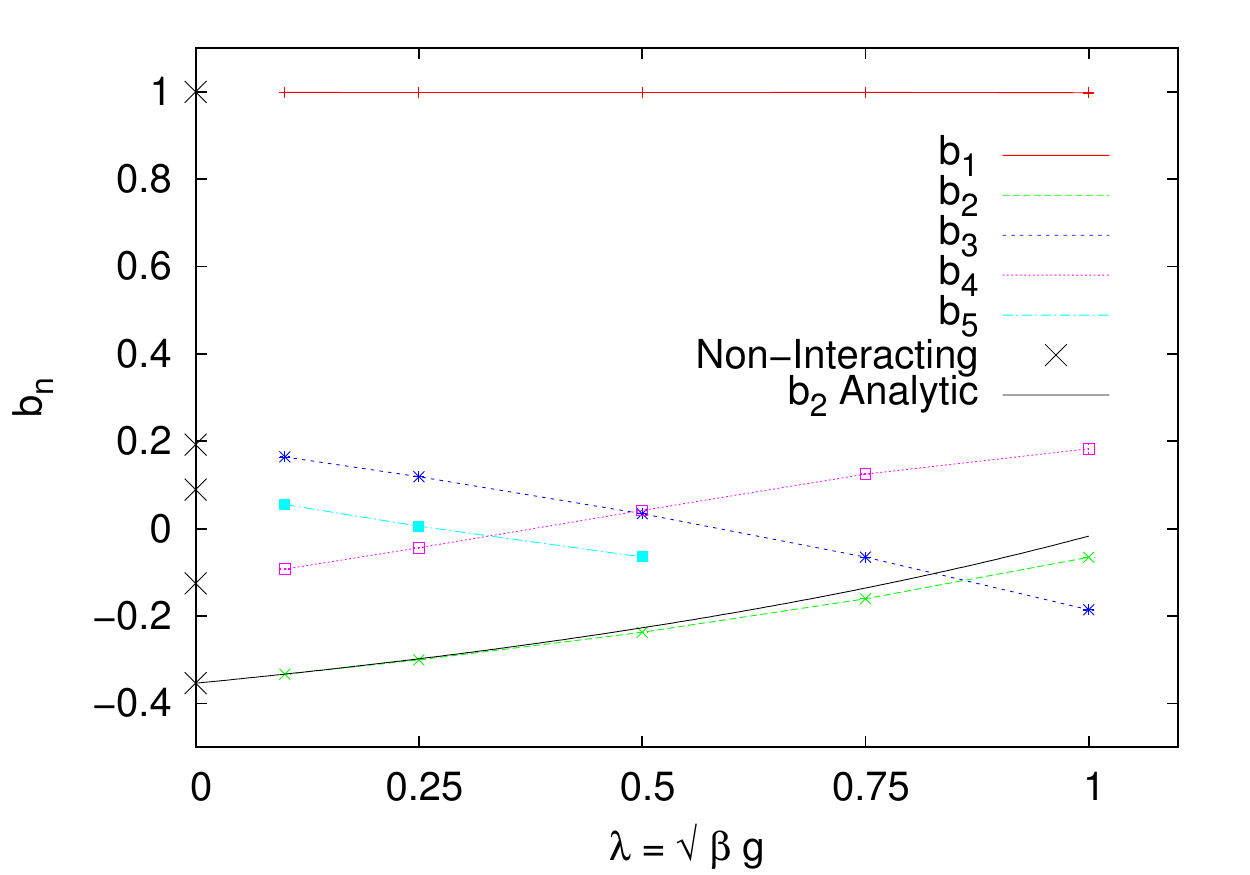}
  \end{center}
  \caption{
  Left panel: Ground-state energy of 10 mass-imbalanced fermions in 1D, as a function of the coupling 
  $\gamma$~\cite{Lukas1}. Dashed lines show a first-order perturbative result.
  Right panel: First few virial coefficients of the attractive 1D Fermi gas, obtained by Fourier-based 
  particle projection.
  }
  \label{Fig:VirialProgress}
\end{figure}

The realization of ultracold atomic gases with different atomic species yields another
area of research where the sign problem prevents useful calculations. However, by implementing imaginary
mass-imbalance and CL approaches~\cite{RBD,BDR,Lukas1}, we have shown that this field is entirely 
open for attractive interactions and at least partially open for the repulsive case. Indeed,
a signal-to-noise issue reappears due to vanishing weight $P[\sigma]$ in the complex $\sigma$ plane
(i.e. the appearance of singularities in $S[\sigma]$) at large repulsive interactions. This issue is currently
open and under investigation. Progress in this direction is shown in the left panel of Fig.~\ref{Fig:VirialProgress},
where we show the ground-state energy (in units of its non-interacting counterpart) of 10 fermions, as a function of the coupling $\gamma$.
The plot shows results for several mass imbalances $\bar m = (m_\uparrow - m_\downarrow)/(m_\uparrow + m_\downarrow)$.
%

\subsection{Particle-projection approach to clusters and high-order virial coefficients}

The idea of implementing a Fourier-transform particle-number projection of a grand-canonical calculation 
to obtain information about finite systems is not new; it is quite common in the context of
nuclear physics (see e.g.~\cite{RMPBender}). However, to our knowledge it has not been applied by way 
of the complex Langevin algorithm, nor has it been applied to the stochastic extraction of virial coefficients
from grand canonical calculations.
Here, we report briefly on progress toward that goal, focusing on 1D systems as test cases, as
done in previous subsections. In Fig.~\ref{Fig:VirialProgress} we show our first results, which indicate
that it may be possible to extract up to fifth order coefficients. In that figure we map out the
coupling-constant dependence of those virial coefficients and compare with the analytic result of
Ref.~\cite{EoS1D} for $b_2$. The remarkable agreement in the case of $b_2$ (up to systematic effects 
as $\lambda$ increases), along with the relative smoothness of the results
for the other coefficients, indicate that the method is robust, although certainly not as precise
as other approaches based on the exact solution of the $n$-body spectrum. Note that the approach
also correctly reproduces $b_1 = 1$ for all couplings. It remains an open question whether the method 
remains viable in higher dimensions, but a priori nothing indicates otherwise.

\section{Conclusions}

In this contribution we have reviewed the recent exploration of the application of complex stochastic 
quantization to the non-relativistic quantum many-body problem. Remarkably, a wide range of applicability 
was found and validated in several ways, e.g. with third-order perturbation theory or imaginary-asymmetry methods.
Specifically, we were able to obtain equations of state of polarized and mass-imbalanced matter with
attractive and repulsive interactions in 1D. We have also shown that practical calculations of high-order
virial coefficients is possible using the idea or particle-number projection along with CL.

\ack
I would like to thank J. Braun, A. C. Loheac, L. Rammelm\"uller, and C. R. Shill for 
discussions and data. This material is based upon work supported by the
National Science Foundation under Grants No. PHY{1452635} (Computational Physics Program).
Numerical calculations have partially been performed at the LOEWE-CSC Frankfurt.


\end{document}